\documentclass[useAMS,onecolumn,usenatbib]{mn2e}
\newif\ifAMStwofonts

\def\pmb#1{\mbox{\boldmath$#1$}}
\def\gtsim {>\kern-1.2em\lower1.1ex\hbox{$\sim$}}
\def\ltsim {<\kern-1.2em\lower1.1ex\hbox{$\sim$}}
\def\gtsim {>\kern-1.2em\lower1.1ex\hbox{$\sim$}}
\def\ltsim {<\kern-1.2em\lower1.1ex\hbox{$\sim$}}

\def\be{\begin{equation}}
\def\ee{\end{equation}}

\def\pmbmt#1{\pmb{\sf #1}}
\def\rmi{{\rm i}}

\usepackage{graphicx}
\usepackage{epsfig}
\begin{document}

\title{Amplitudes of low frequency modes in rotating B type stars}
\author[U. Lee]{ Umin Lee$^1$\thanks{E-mail: lee@astr.tohoku.ac.jp}
\\$^1$Astronomical Institute, Tohoku University, Sendai, Miyagi 980-8578, Japan}

\date{Typeset \today ; Received / Accepted}
\maketitle


\begin{abstract}
Using weakly non-linear theory of oscillation,
we estimate the amplitudes of low frequency modes in a slowly pulsating B (SPB) star,
taking account of the effects of rotation on the modes.
Applying the formulation by Schenk et al (2002), 
we compute non-linear coupling coefficient between the low frequency modes and estimate the equilibrium amplitudes of the modes excited in the star, assuming
the amplitudes of the unstable modes are saturated as a result of non-linear coupling with
stable modes, that is, as a result of parametric instability expected between 
one unstable mode and two stable modes.
We use the traditional approximation to calculate adiabatic and non-adiabatic
oscillations in a rotating star.
We find $r$-modes in a rapidly rotating star play a significant role in the amplitude determination through non-linear coupling.
We also find that for low $m$ modes, the fractional amplitudes of the radiative luminosity caused by the low frequency modes are of order $10^{-4}$ to $10^{-3}$ at the surface.

\end{abstract}


\section{Introduction}

Slowly pulsating B (SPB) stars are a pulsator of low frequency modes excited by
the $\kappa$ mechanism associated with the iron opacity bump at $T\sim 2\times10^5$K
in the interior (e.g., Dziembowski, Moskalik, Pamyatnykh 1993; Gautschy \& Saio 1993).
The observed amplitudes of the oscillations range from $\sim 0.1$ to $\sim1$ mmag (e.g., Huat et al 2009; Diago et al 2009;
Neiner et al 2009; Cameron et al 2008; Balona et al 2011).
Although SPB stars are not necessarily a rapid rotator, 
the effects of rotation on the low frequency modes can be significant, particularly for
$|2\Omega/\omega|\gtsim 1$, where $\Omega$ denotes the rotation frequency and $\omega$
is the oscillation frequency observed in the co-rotating frame of the stars.

Here, we are interested in theoretically determining the amplitudes of low frequency modes in SPB stars,
taking account of the effects of rotation on the modes.
Fully hydrodynamical calculation of radial pulsation has a long history, but
that of non-radial pulsation is not always feasible to carry out, 
since it usually requires a huge amount of numerical 
resources, particularly for low frequency modes in a rotating star.
Instead of hydrodynamical calculation, 
we may apply weakly non-linear theory of 
oscillation to non-radial pulsation, expecting the amplitudes of oscillation modes are 
limited by weak non-linear coupling between them.
As the simplest case,
we may consider non-linear coupling between three modes, whose
amplitudes are expected to reach an equilibrium state as a result of 
parametric instability between one unstable mode
and two stable modes (e.g., Dziembowski 1982; Kumar \& Goldreich 1989;
Wu \& Goldreich 2001; see also Craik 1985).

For the weakly non-linear theory of oscillation, it is essential to calculate the coupling coefficient
between oscillation modes as well as their excitation and damping rates.
To calculate the non-linear coupling coefficient between three oscillation modes, 
we employ the formulation by Schenk et al (2002), who extended the theory to the case of rotating stars.
To compute oscillation modes in a rotating star,
we use the method of calculation by Townsend (2005), who applied the traditional approximation
to calculate both adiabatic and non-adiabatic oscillations of the star.
The traditional approximation is quite helpful to largely reduce the computing time spent for calculating  a large number of coupling coefficients
between various non-radial modes in a rotating star.
We briefly describe the method of solution in \S 2, and \S 3 is for numerical results,
and \S 4 is for conclusions.

\section{method of solution}

\subsection{Linear Oscillation Equation}

Adiabatic oscillation in a uniformly rotating star may be governed by the linear differential equation:
\be
-\omega^2\pmb{\xi}+i\omega\pmbmt{B}\left(\pmb{\xi}\right)+\pmbmt{C}\left(\pmb{\xi}\right)=0,
\ee
where $\omega$ is the oscillation frequency in the co-rotating frame, $\pmb{\xi}$ is the displacement vector,
$
\pmbmt{B}\left({\pmb{\xi}}\right)=2\pmb{\Omega}\times{\pmb{\xi}}
$
with $\pmb{\Omega}$ being the angular velocity vector of rotation, and $\pmbmt{C}$ is the differential operation on $\pmb{\xi}$ and its expression as well as its
derivation may be found, for example, in Schenk et al (2002).
Assuming the time dependence of the perturbations is given by the factor $e^{\rmi\omega t}$, we may write the displacement vector as
\be
\pmb{\xi}(\pmb{x},t)=e^{\rmi\omega t}\pmb{\xi}(\pmb{x})=e^{\rmi\omega t}\left(\xi^r\pmb{e}_r+\xi^\theta\pmb{e}_\theta+\xi^\phi\pmb{e}_\phi\right),
\ee
where $\pmb{e}_r$, $\pmb{e}_\theta$, and $\pmb{e}_\phi$ are the orthonormal base vectors in spherical polar coordinates $(r,\theta,\phi)$. 
We assume that the equilibrium of the rotating star is axisymmetric about the rotation axis,
and that the dependence of the oscillation modes on the azimuthal angle $\phi$ is given by
the factor $e^{\rmi m\phi}$ with $m$ being an integer representing the azimuthal wave number.
The oscillation frequency $\omega$ may be given as $\omega=\sigma+m\Omega$ with $\sigma$ being the oscillation
frequency observed in an inertial frame.
Because of the dependence given by $e^{\rmi(m\phi+\omega t)}$, the oscillation mode with
$m\omega<0$ ($m\omega>0$) is a prograde (retrograde) mode.

Although a separation of variables is in general impossible for oscillations in a rotating star,
it becomes possible under the traditional approximation, in which the term $-\sin\theta\Omega\pmb{e}_\theta$
in $\pmb{\Omega}=\cos\theta\Omega\pmb{e}_r-\sin\theta\Omega\pmb{e}_\theta$ is ignored (e.g., Lee \& Saio 1997).
In the traditional approximation, the components of $\pmb{\xi}(\pmb{x})$ are given by
\be
\xi^r=\xi^r(r)\Theta_{km}(\mu;\nu)e^{\rmi m\phi},
\ee
\be
\xi^\theta={1\over r\omega^2}{p^\prime(r)\over \rho(r)}\Theta^\theta_{km}(\mu;\nu)e^{\rmi m\phi},
\ee
\be
\xi^\phi={1\over r\omega^2}{p^\prime(r)\over \rho(r)}\rmi\Theta^\phi_{km}(\mu;\nu)e^{\rmi m\phi},
\ee
and $p^\prime$ and $\rho^\prime$, which respectively stand for the Eulerian perturbation of the pressure and the density, 
are given by
\be
p^\prime=p^\prime(r)\Theta_{km}(\mu;\nu)e^{\rmi m\phi}e^{\rmi\omega t},
\ee
\be
\rho^\prime=\rho^\prime(r)\Theta_{km}(\mu;\nu)e^{\rmi m\phi}e^{\rmi\omega t},
\ee
where $\mu=\cos\theta$, $\nu=2\Omega/\omega$, $k$ is an integer used as a modal index, and 
\be
\Theta^\theta_{km}(\mu;\nu)={1\over (1-\nu^2\mu^2)\sqrt{1-\mu^2}}\left[-(1-\mu^2){d\over d\mu}+m\nu\mu\right]\Theta_{km}(\mu;\nu),
\ee
\be
\Theta^\phi_{km}(\mu;\nu)={1\over (1-\nu^2\mu^2)\sqrt{1-\mu^2}}\left[-\nu\mu(1-\mu^2){d\over d\mu}+m\right]\Theta_{km}(\mu;\nu).
\ee
The function $\Theta_{km}(\mu;\nu)$, called the Hough function (e.g., Lindzen \& Holton 1968), is the eigenfunction, associated with the eigenvalue $\lambda_{km}$, 
of Laplace tidal equation given by
\be
{\cal L}_\nu\left[\Theta_{km}(\mu;\nu)\right]=-\lambda_{km}\Theta_{km}(\mu;\nu),
\ee
where the definition of the differential operator $\cal L_\nu$ may be found, e.g., in Lee \& Saio (1997).
Note that the oscillation modes in a rotating star are separated into even modes and odd modes,
depending on symmetry of the eigenfunctions about the equator of the star.
For example, the angular dependence of $p^\prime(r,\theta,\phi,t)$ is symmetric (antisymmetric) about the equator
for even (odd) modes.
In this paper, we normalize the function $\Theta_{km}$ as
\be
\int_0^\pi d\theta\int_0^{2\pi}d\phi \sin\theta \left|\tilde\Theta_{km}\right|^2=2\pi\int_{-1}^1d\mu\left|\Theta_{km}\right|^2=1,
\ee
where
\be
\tilde\Theta_{km}=\Theta_{km}e^{\rmi m\phi}.
\ee

For a given azimuthal wavenumber $m$, $\lambda_{km}$ 
depends on the parameter $\nu$
and tends to $l_k(l_k+1)$ with $l_k=|m|+k$ as $\nu\rightarrow 0$ for $k\ge0$,
which corresponds to $\tilde\Theta_{km}\rightarrow Y_{l_k}^m$ as $\nu\rightarrow 0$.
The quantity  $\sqrt{\lambda_{km}}$ represents a kind of surface wave number.
Except for the prograde sectoral modes ($k=0$), $\lambda_{km}$ increases 
as $\nu$ increases.
The prograde sectoral modes (associated with $\lambda_{0m}$ for modes with $m\omega<0$) 
are special modes in rapidly rotating stars whose
surface wavenumber is lower than the value at $\Omega=0$, and hardly changes with $\Omega$. 
Note that $g$-modes belong to $\lambda_{km}$ with positive $k$.
On the other hand, $r$-modes, which are a retrograde mode, belong to $\lambda_{km}$ with
negative $k$ (Lee \& Saio 1997).
In the limit of $\Omega\rightarrow 0$, we have  
$
\omega\rightarrow {2m\Omega/ l_k^\prime(l_k^\prime+1)},
$
where
$l_k^\prime=|m|+|k+1|$ for negative integer $k$.
Note that $\lambda_{km}\rightarrow 0$ as $\omega\rightarrow 2m\Omega/l_k^\prime(l_k^\prime+1)$.

In this paper, we employ the labeling $(l_k,m)$ with $l_k= |m| + k$ for $g$-modes
and the labeling $(l_k^\prime,m)$ with $l_k^\prime=|m|+|k+1|$ for $r$-modes where $k$ is non-negative integer for the former and negative integer for the latter, extending
the familiar notation for non-radial pulsations of a non-rotating star. 
Note that in our convention we have ${\rm mod}(l_k-|m|,2)=0$ and ${\rm mod}(l^\prime_k-|m|,2)=1$
for even modes, while ${\rm mod}(l_k-|m|,2)=1$ and ${\rm mod}(l^\prime_k-|m|,2)=0$ for odd modes.

\subsection{Weakly Nonlinear Oscillation Equation and Parametric Instability}

Nonlinear evolution of small amplitude oscillation modes in a uniformly rotating star is governed by 
the oscillation equation with nonlinear terms:
\be
\ddot{\pmb{\xi}}+\pmbmt{B}\left(\dot{\pmb{\xi}}\right)+\pmbmt{C}\left(\pmb{\xi}\right)=
\pmb{a}^{(2)}\left(\pmb{\xi},\pmb{\xi}\right),
\ee
where $\dot{\pmb{\xi}}=d\pmb{\xi}/dt$ and $\ddot{\pmb{\xi}}=d^2\pmb{\xi}/dt^2$, 
and $\pmb{a}^{(2)}\left(\pmb{\xi},\pmb{\xi}\right)$ represents a collection of nonlinear terms of second order in $\pmb{\xi}$, and the $i$th component of $\pmb{a}^{(2)}$ is given by (Schenk et al 2002)
\be
a_i^{(2)}\left(\pmb{\xi},\pmb{\xi}\right)=-{\rho^{-1}}\nabla_j\left\{p\left[\left(\Gamma_1-1\right)\Pi_i^j+\Xi_i^j
+\Psi\delta_i^j\right]\right\}-({1/ 2})\xi^k\xi^l\nabla_k\nabla_l\nabla_i\Phi,
\ee
where $\nabla_j$ denotes the covariant derivative with respect to the coordinate $x^j$, 
\be
\Pi_i^j=(\nabla_i\xi^j)\nabla\cdot\pmb{\xi},
\ee
\be
\Xi_i^j=(\nabla_i\xi^k)(\nabla_k\xi^j), 
\ee
\be
\Psi=\left({1/ 2}\right)\Pi\left[\left(\Gamma_1-1\right)^2+{\partial\Gamma_1/\partial\ln\rho}\right]+\left({1/2}\right)\left(\Gamma_1-1\right)\Xi,
\ee
$
\Pi=\delta^i_j\Pi^j_i=\left(\nabla\cdot\pmb{\xi}\right)^2,  
$
$
\Xi=\delta^i_j\Xi^j_i=\left(\nabla_j\xi^k\right)\left(\nabla_k\xi^j\right),
$
$\delta^i_j$ is the Kronecker delta, $\Phi$ is the gravitational potential,
and the repeated indices imply 
the summation over the indices from 1 to 3, and $\Gamma_1=\left(\partial\ln p/\partial\ln \rho\right)_{\rm ad}$.
Note that we have applied the Cowling approximation, neglecting the Eulerian perturbation of 
the gravitational potential.

Following Schenk et al (2002), we use eigenvalues $\omega$ and eigenfunctions $\pmb{\xi}$ of the linear oscillation equation (1) to
expand the displacement vector $\pmb{\xi}(\pmb{x},t)$ and its time derivative $\dot{\pmb{\xi}}(\pmb{x},t)$ in the nonlinear equation (13):
\be
\left[\matrix{\pmb{\xi}(\pmb{x},t)\cr\dot{\pmb{\xi}}(\pmb{x},t)\cr}\right]=\sum_Ac_A(t)\left[\matrix{\pmb{\xi}_A(\pmb{x})\cr \rmi\omega_A\pmb{\xi}_A(\pmb{x})\cr}\right],
\ee
for which
\be 
\sum_A\left(\dot c_A-\rmi\omega_A c_A\right)\pmb{\xi}_A(\pmb{x})=0,
\ee
where the subscript $A$ stands for a collection of numbers such as harmonic degree $l$, azimuthal order $m$, and
radial order $n$ used to identify a linear mode.
Note that if a mode with with $(m_A,\omega_A)$ satisfies the linear oscillation equation 
the mode with $(-m_A,-\omega_A)$ also satisfies the same equation, and both modes are included in the expansion given above.
Substituting the expansion (18) into the governing equation (13), and
making a scaler product with $\pmb{\xi}_A^*$ and integrating over the volume of the star, we obtain
\be
\dot c_A(t)-\rmi\omega_Ac_A(t)=-{\rmi}\left<\pmb{\xi}_A,\pmb{a}^{(2)}\left(\pmb{\xi},\pmb{\xi}\right)\right>/b_A,
\ee
where
\be
c_A(t)=\left<\pmb{\xi}_A,\omega_A\pmb{\xi}(t)-\rmi\dot{\pmb{\xi}}(t)-\rmi\pmbmt{B}\left(\pmb{\xi}(t)\right)\right>/b_A,
\ee
\be
b_A=-\left<\pmb{\xi}_A,\rmi\pmbmt{B}\left(\pmb{\xi}_A\right)\right>+2\omega_A\left<\pmb{\xi}_A,\pmb{\xi}_A\right>,
\ee
and for $A\not= B$ we have used a modified type of orthogonality relation given by
\be
\left<\pmb{\xi}_A,\rmi\pmbmt{B}\left(\pmb{\xi}_B\right)\right>-\left(\omega_A+\omega_B\right)\left<\pmb{\xi}_A,
\pmb{\xi}_B\right>=0,
\ee
where
\be
\left<\pmb{\xi}_A,\pmb{\xi}_B\right>=\int d^3\pmb{x}\rho(\pmb{x})\pmb{\xi}^*_A(\pmb{x})\cdot\pmb{\xi}_B(\pmb{x}),
\ee
and the asterisk ${}^*$ in the superscript implies the complex conjugation.
If we substitute the expansion
$
\pmb{\xi}\left(\pmb{x},t\right)=\sum_Bc_B^*(t)
\pmb{\xi}_B^*\left(\pmb{x}\right)
$
into $\pmb{a}^{(2)}\left(\pmb{\xi},\pmb{\xi}\right)$, we may obtain
\be
\dot c_A(t)-\rmi\omega_Ac_A(t)=-\rmi\omega_A\sum_{B,C}\left({\kappa^*_{ABC}/\epsilon_A}\right)c_B^*(t)c_C^*(t),
\ee
where
\be
\kappa_{ABC}=\left<\pmb{\xi}_A^*,\pmb{a}^{(2)}\left(\pmb{\xi}_B,\pmb{\xi}_C\right)\right>, 
\ee
and $\epsilon_A\equiv \omega_A b_A$ corresponds to the total energy of oscillation in the co-rotating frame of the star (e.g., Lee \& Saio 1990).
If we introduce
$
{\hat c}_A=c_A\exp(-\rmi \omega_A t),
$
the equation (25) reduces to
\be
\dot{\hat c}_A=-\rmi \omega_A
\sum_{B,C}\left({\kappa^*_{ABC}/ \epsilon_A}\right){\hat c}_B^*(t){\hat c}_C^*(t)e^{-\rmi\Delta\omega t},
\ee
where
$
\Delta\omega=\omega_A+\omega_B+\omega_C.
$

If the driving rate of an unstable mode is smaller than the damping rates of stable modes nonlinearly coupled with the unstable one,
the growth of the unstable mode can be saturated by a transfer of energy to a small number of the damped modes.
Here, we consider parametric instability between three modes, one unstable mode and two stable modes and
we call the former the parent mode and the latter the daughter modes, and we expect the amplitude of the parent mode
is saturated by energy transfer to the daughter modes.
In the following, for convenience, we call mode $A$ the parent and modes $B$ and $C$ the daughter modes.
If we consider nonlinear mode coupling between three modes $A$, $B$, and $C$, we obtain
\be
\dot{\hat c}_A=-\gamma_A\hat c_A-\rmi \omega_A
{\eta^*_{ABC}}{\hat c}_B^*(t){\hat c}_C^*(t)e^{-\rmi\Delta\omega t},
\ee
and two similar  equations for $\dot{\hat c}_B$ and $\dot{\hat c}_C$,
where we have included the effects of linear destabilization ($\gamma<0$) and stabilization ($\gamma>0$) of the modes, and 
we have normalized the eigenfunctions $\pmb{\xi}_A$, $\pmb{\xi}_B$, and $\pmb{\xi}_C$ such that
$\epsilon_A=\epsilon_B=\epsilon_C=GM^2/R$, which leads to
$
{\kappa_{ABC}/\epsilon_A}={\kappa_{BCA}/\epsilon_B}={\kappa_{CAB}/\epsilon_C}\equiv{\eta_{ABC}/ 2}.
$

Parametric instability may occur when the amplitude of the parent mode, $|c_A|$, exceeds
the critical amplitude given by (e.g., Dziembowski 1982; Arras et al 2003)
\be
|c_{A:c}|^2={1\over |\eta_{ABC}|^2Q_BQ_C}\left[1+\left({\Delta\omega\over\gamma_B+\gamma_C}\right)^2\right],
\ee
where $Q_j=-\omega_j/\gamma_j$.
The equilibrium amplitude of the parent mode is then given by
\be
|c_{A:e}|^2={1\over|\eta_{ABC}|^2Q_BQ_C}\left[1+\left({\Delta \omega\over\Delta\gamma}\right)^2\right],
\ee
and those of the daughter modes are by
\be
\left|{c_{B:e}}\right|^2=\left|c_{A:e}\right|^2{Q_B/ Q_A}, \quad
\left|{c_{C:e}}\right|^2=\left|c_{A:e}\right|^2{Q_C/ Q_A},
\ee
where $\Delta\gamma=\gamma_A+\gamma_B+\gamma_C$.
Here, we have assumed $Q_BQ_C>0$, $Q_CQ_A>0$, and $Q_AQ_B>0$, which is equivalent to the relation given by
$Q_A>0,~Q_B>0,~Q_C>0$ or by $Q_A<0,~Q_B<0,~Q_C<0$, that is, the signs of 
$\omega_B$ and $\omega_C$ are the same to each other but are different from that of $\omega_A$, because
the parent mode $A$ is assumed unstable ($\gamma_A<0$) and the daughter modes $B$ and $C$
stable ($\gamma_B>0$ and $\gamma_C>0$).
Since $\omega_B$ and $\omega_C$ have the same sign, to obtain a resonant coupling satisfying 
$\Delta\omega\sim0$, we have
$|\omega_B|\ltsim ~|\omega_A|$ and $|\omega_C|\ltsim ~|\omega_A|$.
We use the condition $\Delta\gamma>0$ as the criteria for effectively stable equilibrium state 
of three mode coupling (e.g., Wu \& Goldreich 2001; Arras et al 2003).

One of the selection rules giving non-zero coupling coefficient $\eta_{ABC}\not=0$ is 
\be
m_A+m_B+m_C=0,
\ee
and another selection rule may be simply stated that
the coupling coefficient $\eta_{ABC}$ is non-zero only when the mode triad is composed of
three even modes or of one even mode and two odd modes (e.g., Schenk et al 2002).
For $m_A<0$, for example, we have two cases because of the selection rule (32), that is, both $m_B$ and $m_C$ are positive or one of $m_B$ and $m_C$ is negative so that $m_Bm_C<0$.
In the former case, if the parent mode is a prograde (retrograde) mode, the two daughter modes are
prograde (retrograde) modes.
In the latter case, however, if the parent mode is a prograde mode having $\omega_A>0$, 
the daughter mode with $m<0$ is a retrograde mode since $\omega_B<0$ and $\omega_C<0$.
On the other hand, if the parent mode is a retrograde mode having $\omega_A<0$, 
the daughter mode with $m<0$ is a prograde mode since $\omega_B>0$ and $\omega_C>0$.

\section{Numerical Results}

As a background model for mode calculation, we use a $4M_\odot$ main sequence model
computed by a standard stellar evolution code, 
where we have used OPAL opacity (Iglesias \& Rogers 1996).
The physical parameters of the model are $\log T_{\rm eff}=4.142$, $\log (L/L_\odot)=2.470$, $R/R_\odot=2.980$, and $X_c=0.4602$, 
and the initial abundance is given by $X=0.7$ and $Z=0.02$, where $T_{\rm eff}$, $L$, $R$, and $X_c$ respectively denote
the effective temperature, the surface luminosity, the radius of the model, and
the hydrogen mass fraction at the center.
Since we use the main sequence model slightly evolved from ZAMS, the model have a thin $\mu$-gradient
zone above the convective core where $\mu$ denotes the mean molecular weight.
We employ the method of calculation given by Townsend (2005) to calculate in the traditional approximation
both adiabatic and non-adiabatic modes of a uniformly rotating star, where
no effects of rotational deformation are considered.
We also employ the Cowling approximation, which is good enough
for high radial order $g$-modes of low degree $l$.
For this model, as well as pulsationally stable low frequency modes, 
we obtain many unstable $g$-modes and $r$-modes, which are excited by the $\kappa$ mechanism
associated with the iron opacity bump located at $T\sim 2\times10^5$K.
It is our main concern here how non-linear three mode coupling determines the amplitudes
of the low frequency modes in a rotating B-type star.
We use the eigenfrequencies and eigenfunctions of adiabatic modes to compute
the nonlinear coupling coefficient $\eta_{ABC}$.
The excitation and damping rates
$\gamma$ are given by the imaginary part of
the complex eigenfrequency, $\omega_{\rm I}={\rm Im}(\omega)$, which is obtained by non-adiabatic mode calculation.

In this paper, to prepare a set of daughter modes used to calculate the coupling coefficient $\eta_{ABC}$ for a given low $|m|$ parent mode, 
we consider low frequency modes of $|m|$ ranging from $|m|=0$ to 5 and of
$l$ in the limited range of $|m|\le l\le |m|+1$ 
for $|m|\not=0$ and $l=1$ and 2 for $m=0$, that is, we compute
even and odd $g$-modes with $(l,m)=(|m|,m)$ and $(|m|+1,m)$, and 
odd and even $r$-modes with $(l^\prime,m)=(|m|,m)$ and $(|m|+1,m)$ 
for $|m|=1$ to 5, and even and odd $g$-modes with $(l,m)=(2,0)$ and $(1,0)$ for $m=0$, in the frequency range
$0.05\le\bar\omega\le 2$, where $\bar\omega\equiv\omega/\sqrt{GM/R^3}$, and 
$M$ is the mass of the star and $G$ is the gravitational constant.
Note that the $r$-modes are in the frequency range of $|\omega|<2|m|\Omega/l^\prime(l^\prime+1)$.
For a given combination of $(m_A,m_B,m_C)$ for mode triad, 
the even/odd mode combinations giving non-zero $\eta_{ABC}$ are 
$(A_e,B_e,C_e)$, $(A_e,B_o,C_o)$, $(A_o,B_e,C_o)$, and
$(A_o,B_o,C_e)$, where the subscripts $e$ and $o$ stand for even and odd modes, respectively.

For a given unstable parent mode, there are numerous combinations of a pair of stable daughter modes,
satisfying the selection rules for non-zero coupling coefficient $\eta_{ABC}$.
For each of the combinations we 
compute the critical amplitude $c_{A:c}$ and equilibrium amplitude $c_{A:e}$ for the parent mode using equations (29) and (30),
where the excitation and damping rates are obtained by non-adiabatic mode calculation.
Among the critical amplitudes calculated for various combinations of a pair of daughter modes for a given parent mode, we chose the smallest one,
considering that the parent mode reaches the smallest critical amplitude first to be in an equilibrium state.
Since we search for the combination giving the smallest $c_{A:c}$ from a limited set of daughter modes for a parent mode,
the critical amplitude $c_{A:c}$ thus determined should be regarded as an upper limit for the parent mode.
As indicated by equation (29),
the critical amplitude $|c_{A:c}|$ becomes smaller for larger values of $|\eta_{ABC}|$ and $\sqrt{Q_BQ_C}$ 
and has a dip at $\Delta \omega=0$ because of the factor $1+[\Delta\omega/(\gamma_B+\gamma_C)]^2$ in equation (29).
Since normalized damping rates $\bar\gamma\equiv\gamma/\sqrt{GM/R^3}$ of high radial order $g$-modes of the model
are of order $10^{-5}\sim 10^{-4}$, which are much larger than the growth rates ranging from
$|\bar\gamma|\sim10^{-8}$ to $\sim 10^{-5}$ for the $g$-modes,
the dip at $\Delta\omega=0$ will be very sharp for triads of $g$-modes having frequencies $|\bar\omega|\gtsim 0.05$.
In this paper, 
we search for the smallest critical amplitude $c_{A:c}$ for a parent mode
among mode triads satisfying $\Delta\omega\sim 0$, that is, nearly in frequency resonance.
This procedure is helpful to 
reduce the number of combinations of daughter modes we have to try in order to find the smallest $|c_{A:c}|$.
Because of the assumption $\Delta\omega\sim0$, the factor
$1+[\Delta\omega/(\gamma_B+\gamma_C)]^2$ in equation (29) takes values between $\sim 1$ and $\sim10$ for most of the triads examined, 
and $|c_{A:c}|$ is practically dependent on the two quantities $|\eta_{ABC}|$ and $\sqrt{Q_BQ_C}$, that is,
the smallest critical amplitude is likely to take place when either $|\eta_{ABC}|$ or $\sqrt{Q_BQ_C}$ is very large or when both of them are large.
Since $\eta_{ABC}$, which is calculated by using the eigenfunctions of adiabatic modes, 
is the sum of products of the three eigenfunctions of
modes $A$, $B$, and $C$, and since the eigenfunction of $g$-modes has an asymptotic form proportional to
$\cos\left(\int^r k_{r}dr\right)$ with $k_r$ being the wave number in the radial direction and
$\int k_{r}dr=n_{g}\pi$ with $n_g$ being the number of $g$-nodes of the eigenfunction
when integrated over the entire propagation zone of the $g$-modes,
the sum of terms proportional to $I_{ABC}=\int_0^R dr f(r)\cos\left(\int^r k_{rA}dr\right)\cos\left(\int^r k_{rB}dr\right)\cos\left(\int^r k_{rC}dr\right)$ with $f(r)$ being a spatially slowly varying weighting function may be maximized when
$n_{gA}\sim|n_{gB}-n_{gC}|$, which is 
numerically confirmed (see also Wu \& Goldreich 2001).
The quantity $\sqrt{Q_BQ_C}$ can be large when
a low radial order $g$-mode or $r$-mode with a very small damping rate is in the mode triad, and in this case
we do not necessarily have the property $n_{gA}\sim|n_{gB}-n_{gC}|$.

\subsection{Slow Rotation}

\begin{figure}
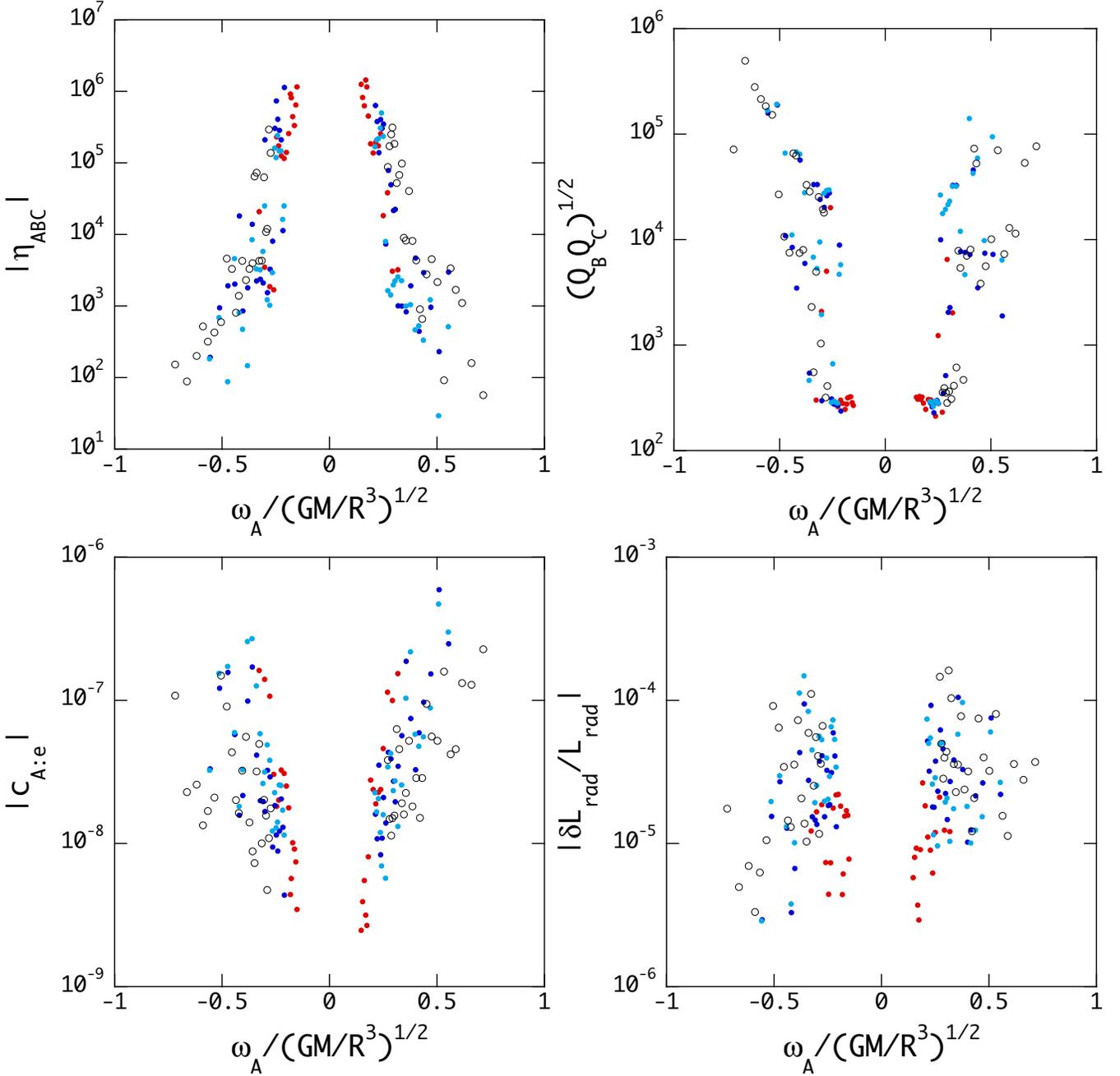

\resizebox{0.5\columnwidth}{!}{
\includegraphics{f1a.epsi}}
\resizebox{0.5\columnwidth}{!}{
\includegraphics{f1b.epsi}}
\resizebox{0.5\columnwidth}{!}{
\includegraphics{f1c.epsi}}
\resizebox{0.5\columnwidth}{!}{
\includegraphics{f1d.epsi}}
\caption{Mode triad quantities $|\eta_{ABC}|$, $\sqrt{Q_BQ_C}$, $|c_{A:e}|$, and $|\delta L_{\rm rad}/L_{\rm rad}|$ are plotted
versus the frequency $\omega_A/\sqrt{GM/R^3}$ of the parent mode for a $4M_\odot$ main sequence
model
for $\Omega/\sqrt{GM/R^3}=0.01$, where 
$\delta L_{\rm rad}$ denotes the Lagrangian variation of 
the surface luminosity $L_{\rm rad}$ caused by the parent mode, $Q_BQ_C =\omega_B\omega_C/\gamma_B\gamma_C$, and
the red, blue, cyan dots and black open circle stand for the parent modes of 
$(l_A,m_A)=(1,-1)$, $(2,-1)$, $(2,-2)$, and $(3,-2)$, respectively.
Note that for a given parent mode a combination of a pair of daughter modes has been chosen so that the critical amplitude $|c_{A:c}|$ of the parent mode
be smallest, and that the thus determined combination of daughter modes makes the mode triad associated with the parent mode.
Here, the parent modes with positive (negative) $\omega_A$ are prograde (retrograde) modes.
}
\end{figure}

\begin{figure}
\resizebox{0.33\columnwidth}{!}{
\includegraphics{f2a.epsi}}
\resizebox{0.33\columnwidth}{!}{
\includegraphics{f2b.epsi}}
\resizebox{0.33\columnwidth}{!}{
\includegraphics{f2c.epsi}}
\caption{Coupling coefficient $\eta_{ABC}(r)\equiv\kappa_{ABC}(r)/\epsilon_A$ and eigenfunctions $xz_1$ and $xz_2/(c_1\bar\omega^2)$ 
versus $x\equiv r/R$ for a mode triad
composed of a parent mode of $l_A=-m_A=1$ and daughter modes of $l_B=-m_B=4$ and $l_C=m_C=5$ for $\bar\Omega=0.01$, 
where the eigenfunctions $z_1$ and $z_2$ are normalized so that
the oscillation energy $\epsilon_A$ in the co-rotating frame be equal to $GM^2/R$ with $G$, $M$ and $R$ being the gravitational constant, 
and the mass and radius of the star, respectively. 
See the Appendix B for the definition of the quantities $c_1$, $z_1$ and $z_2$.
Here, $\bar\omega_A=0.1680$,
$\bar\omega_B=-0.07082$, and $\bar\omega_C=-0.09769$.
In panels (b) and (c), the black, red, and blue lines indicate the mode A, B, and C, respectively.}
\end{figure}

In Figure 1, the quantities $|\eta_{ABC}|$, $\sqrt{Q_BQ_C}$, $|c_{A:e}|$, 
and $|\delta L_{\rm rad}/L_{\rm rad}|$ computed for the mode triad composed of low frequency $g$-modes and 
corresponding to the smallest $|c_{A:e}|$ are plotted versus the frequency $\omega_A/\sqrt{GM/R^3}$ of the parent mode 
for the case of $\bar\Omega\equiv\Omega/\sqrt{GM/R^3}=0.01$, where
$\delta L_{\rm rad}$ is the Lagrange variation of the radiative luminosity at the surface caused by the parent mode, and 
we have considered only $g$-modes for the mode triads, ignoring $r$-modes because of the slow rotation.
Note that we have $|c_{A:e}|\sim|c_{A:c}|$ for the mode triads plotted in the figure.
Since the ratio $|2\Omega/\omega|$ is much smaller than unity for unstable $g$-modes (parent modes) for the slow rotation, 
the distributions of the points in the figure are almost
symmetric between prograde and retrograde modes, although there exists slight deviation from
the symmetry.
As suggested by the panels (a) to (c), $|c_{A:c}|$ for the 
parent modes can be small either when $|\eta_{ABC}|$ or $\sqrt{Q_BQ_C}$ is very large or when both of them
are large.
For the mode triads in the figure, 
$|\eta_{ABC}|$ ($|c_{A:e}|$) tends to increase (decrease) with decreasing $|\bar\omega|$, but
the dependence of $|\delta L_{\rm rad}/L_{\rm rad}|$ on $\bar\omega$
is not simple because the normalizing amplitudes, defined to make the energy of oscillation equal to $GM^2/R$, increase as $|\bar\omega|$ decreases.
The fractional luminosity amplitudes $|\delta L_{\rm rad}/L_{\rm rad}|$ at the surface 
take values ranging from $\sim 10^{-6}$ to $\sim 10^{-4}$, and those of the parent $l=|m|=1$ $g$-modes tend to be smaller than the others.

As shown by Figure 1, the coupling coefficient $|\eta_{ABC}|$ can be as large as $\sim 10^6$ for the very low frequency parent modes for $\bar\Omega=0.01$.
In Figure 2, the coupling coefficient $\eta_{ABC}(r)=\kappa_{ABC}(r)/\epsilon_A$ (see Appendix B) and the eigenfunctions $x z_1$ and $x z_2/(c_1\bar\omega^2)$
are plotted 
versus $x\equiv r/R$ for a mode triad composed of a parent mode of $l_A=-m_A=1$ and daughter modes of $l_B=-m_B=4$ and $l_C=m_C=5$
for $\bar\Omega=0.01$, where the eigenfunctions are normalized so that the oscillation energy be equal to $GM^2/R$.
The panel (a) shows that the mean magnitude of $|\eta_{ABC}(r)|$ increases rapidly with increasing $r$ in the $\mu$-gradient region above the convective core,
and the panel (c) indicates that this rapid increase is caused by the amplitude trapping of the eigenfunctions into the $\mu$-gradient zone.
Note that $\eta_{ABC}(r)$ stays almost constant outside the $\mu$-gradient region as shown by panel (a).
For this mode triad, the modes $B$ and $C$ are very high radial order $g$-modes, for which
we have $n_{gA}\sim |n_{gB}-n_{gC}|$.
The panel (a) suggests that for the mode triad the non-linear coupling between the modes preferentially occurs in the thin $\mu$-gradient zone
having a high Brunt-V\"ais\"al\"a frequency.
Note that for the ZAMS model with no $\mu$-gradient zone, there occurs no rapid increase in the mean magnitude of $|\eta_{ABC}(r)|$ immediately above the 
convective core.

\begin{figure}
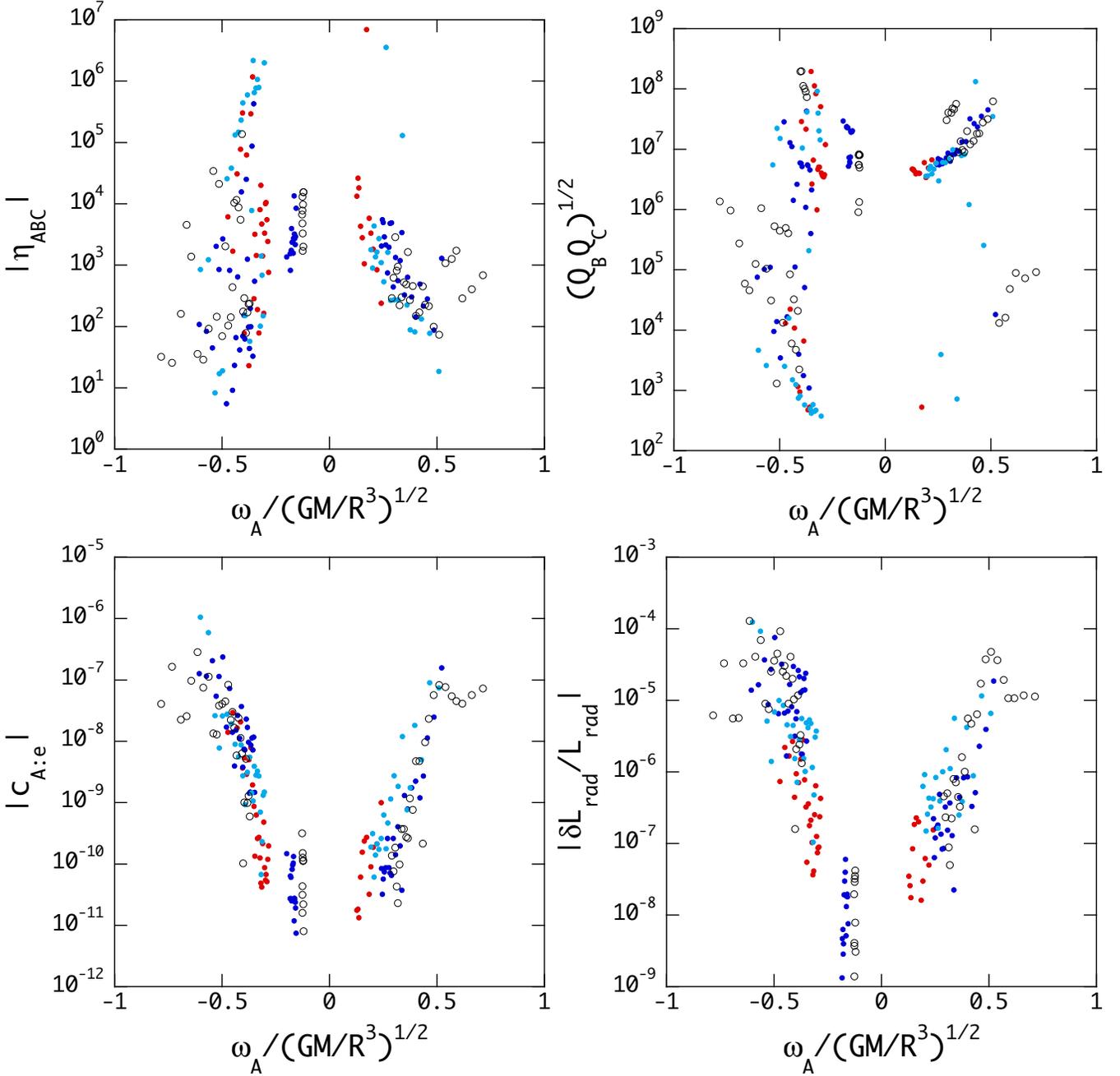

\resizebox{0.5\columnwidth}{!}{
\includegraphics{f3a.epsi}}
\resizebox{0.5\columnwidth}{!}{
\includegraphics{f3b.epsi}}
\resizebox{0.5\columnwidth}{!}{
\includegraphics{f3c.epsi}}
\resizebox{0.5\columnwidth}{!}{
\includegraphics{f3d.epsi}}
\caption{Same as Figure 1 but for the case of $\bar\Omega=0.2$. Here, we have included $r$-modes as a possible member in the mode triads
to calculate the coupling coefficient $\eta_{ABC}$.}
\end{figure}

\subsection{Rapid Rotation}

For weakly non-linear coupling of oscillations in a rapidly rotating star, $r$-modes may come into play as an important member 
in mode triads for parametric instability.
For a rotation rate $\bar\Omega=0.2$, for example,
the maximum frequency $2|m|\bar\Omega/l^\prime(l^\prime+1)$ for even (odd) $r$-modes in the co-rotating frame
is 1/15 (0.2), 1/15 (2/15), 0.06 (0.1), 4/75 (0.08), 1/21 (1/15) for $|m|=1$ to $|m|=5$, and the 
corresponding inertial frame frequency $|\bar\sigma|$ is
2/15 (0), 1/3 (4/15), 0.54 (0.5), 56/75 (0.72), and 20/21 (14/15), respectively.
In Figure 3, the mode triad quantities $|\eta_{ABC}|$, $\sqrt{Q_BQ_C}$, $|c_{A:e}|$, 
and $|\delta L_{\rm rad}/L_{\rm rad}|$ are plotted versus the frequency $\bar\omega_A$ of the low $|m|$ parent modes 
for $\bar\Omega=0.2$.
The distribution of the points of the parent modes of $l=|m|=1$, for example, 
is not symmetric any more between prograde and retrograde modes, because 
the frequency spectra of low frequency $g$-modes themselves largely deviate from the symmetry for $|2\Omega/\omega|\gtsim 1$,
and $r$-modes appear only as a retrograde mode.
Since the damping rates $\bar\gamma$ of low radial order $r$-modes can be smaller than $10^{-10}$
and those of low radial order $g$-modes are of order $10^{-8}$ to $10^{-9}$, it is likely that
the mode triads giving
the smallest critical amplitude $|c_{A:c}|$ for the parent modes are those containing a low radial order $r$-mode or $g$-mode.
Figure 4 shows mode triad quantities $|\eta_{ABC}|$, $\sqrt{Q_BQ_C}$, 
and $|\delta L_{\rm rad}/L_{\rm rad}|$ the same as those plotted in Figure 3, but here
the red, blue, and cyan dots respectively stand for the mode triads containing no $r$-modes, one $r$-mode, 
and two $r$-modes.
The low frequency parent modes tend to be coupled to one $r$-mode or two.
We find cases where the parent modes are an unstable $r$-mode coupled with
a stable $r$-mode.
Since both $|\eta_{ABC}|$ and $\sqrt{Q_BQ_C}$ are large, the amplitude $|\delta L_{\rm rad}/L_{\rm rad}|$
becomes very small.
We also find cases in which the parent modes are a low frequency even $g$-mode coupled with
two stable odd $r$-modes, one low radial order and the other high radial order $r$-modes.
If no $r$-modes are in a mode triad, on the other hand, the amplitude $|\delta L_{\rm rad}/L_{\rm rad}|$ for
the parent mode is in general larger than those of the mode triads that contain one $r$-mode or two.

\begin{figure}
\resizebox{0.33\columnwidth}{!}{
\includegraphics{f4a.epsi}}
\resizebox{0.33\columnwidth}{!}{
\includegraphics{f4b.epsi}}
\resizebox{0.33\columnwidth}{!}{
\includegraphics{f4c.epsi}}
\caption{Mode triad quantities $|\eta_{ABC}|$, $\sqrt{Q_BQ_C}$, and $|\delta L_{\rm rad}/L_{\rm rad}|$ are plotted
versus the frequency $\omega_A/\sqrt{GM/R^3}$ of the parent mode for a $4M_\odot$ main sequence
model for the case of $\bar\Omega=0.2$, where 
$\delta L_{\rm rad}$ denotes the Lagrangian variation of 
the surface luminosity $L_{\rm rad}$ caused by the parent mode, $Q_BQ_C =\omega_B\omega_C/\gamma_B\gamma_C$, and
the red, blue, cyan dots indicate the mode triads containing
no $r$-modes, one $r$-mode, and two $r$-modes, respectively.
Here, the parent modes with positive (negative) $\omega_A$ are prograde (retrograde) modes.}
\end{figure}

In Figure 5, the quantities $\eta_{ABC}(r)$, $x z_1$, and $x z_2/(c_1\bar\omega^2)$ are plotted
versus $x=r/R$ for a mode triad composed of a parent mode of $l_A=-m_A=2$ and daughter modes of $l_B-1=m_B=1$ and $l_C-1=m_C=1$ for $\bar\Omega=0.2$, 
where the eigenfunctions are normalized so that the oscillation energy be equal to $GM^2/R$.
Here, the daughter mode C is an $r$-mode, and the eigenfunction $z_1$ of the $r$-mode has an amplitude much smaller than those of
the $g$-modes A and B, although the amplitude $z_2/(c_1\bar\omega^2)$ of the $r$-mode is
comparable to the $g$-modes.
The figure suggests that the terms containing $z_2/(c_1\bar\omega^2)$ and/or $d[z_2/(c_1\bar\omega^2)]/dx$ make
dominating contributions to $\eta_{ABC}$ (see Appendix B).
We also note that the amplitudes of the modes in the triad are not strongly trapped in the $\mu$-gradient zone.
Although the coefficient $\eta_{ABC}(r)$ is spatially oscillatory 
in the region $0.1\ltsim x\ltsim 0.3$, it takes almost a constant value in $x\gtsim 0.5$.

\begin{figure}
\resizebox{0.33\columnwidth}{!}{
\includegraphics{f5a.epsi}}
\resizebox{0.33\columnwidth}{!}{
\includegraphics{f5b.epsi}}
\resizebox{0.33\columnwidth}{!}{
\includegraphics{f5c.epsi}}
\caption{Coupling coefficient $\eta_{ABC}(r)\equiv\kappa_{ABC}(r)/\epsilon_A$ and eigenfunctions $xz_1$ and $xz_2/(c_1\bar\omega^2)$ 
versus $x\equiv r/R$ for a mode triad
composed of a parent mode of $l_A=-m_A=2$ and daughter modes of $l_B-1=m_B=1$ and $l_C-1=m_C=1$ for $\bar\Omega=0.2$, 
where the eigenfunctions $z_1$ and $z_2$ are normalized so that
the oscillation energy $\epsilon_A$ in the co-rotating frame be equal to $GM^2/R$ with $G$, $M$ and $R$ being the gravitational constant, 
and the mass and radius of the star. 
See the Appendix B for the definition of the quantities $c_1$, $z_1$ and $z_2$.
Here, $\bar\omega_A=-0.5332$,
$\bar\omega_B=0.3329$, and $\bar\omega_C=0.1999$, and the mode C is an $r$-mode.
In panels (b) and (c), the black, red, and blue lines indicate the mode A, B, and C, respectively.}
\end{figure}

\subsection{In an Inertial Frame}

\begin{figure}
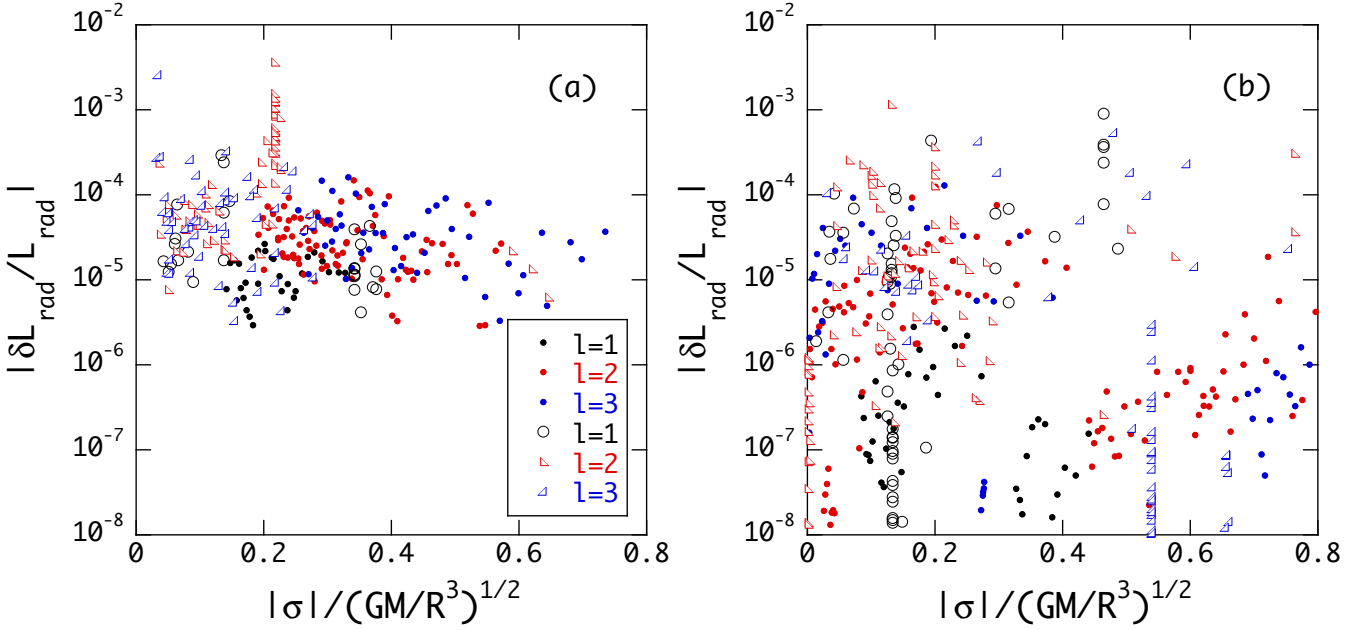

\resizebox{0.5\columnwidth}{!}{
\includegraphics{f6a.epsi}}
\resizebox{0.5\columnwidth}{!}{
\includegraphics{f6b.epsi}}
\caption{Fractional amplitude $|\delta L_{\rm rad}/L_{\rm rad}|$ of the surface luminosity
versus the oscillation frequency $|\sigma|/\sqrt{GM/R^3}$ observed in an inertial frame for
$\bar\Omega=0.01$ in panel (a) and for $\bar\Omega=0.2$ in panel (b), where the filled (open)
symbols stand for the parent (daughter) modes, and the legend of the symbols for both panels are given in
the lower right corner in panel (a), and
only the modes of $1\le l\le3$ are plotted.
For the parent modes, the combination $(l,|m|)$ is given by $(1,1)$, $(2,1)$, $(2,2)$, and $(3,2)$, while
for the daughter modes the combination $(l,|m|)$ is given by $(1,0)$, $(1,1)$, $(2,0)$, $(2,1)$,$(2,2)$, $(3,2)$, and $(3,3)$.
 }
\end{figure}

It may be useful to plot the fractional amplitude of both the parent and daughter modes
as a function of the oscillation frequency
observed in an inertial frame, where the daughter modes are regarded as being non-linearly excited
by the parametric instability. 
Figure 6 shows $|\delta L_{\rm rad}/L_{\rm rad}|$
versus the inertial frame oscillation frequency $|\bar\sigma|$ for the case of
$\bar\Omega=0.01$ (panel a) and $\bar\Omega=0.2$ (panel b), where $\bar\sigma=\bar\omega-m\bar\Omega$,
and the filled (open) symbols stand for the parent (daughter) modes.
Here, only the daughter modes having $l\le 3$ are plotted in the figure.
If one of the daughter modes in a mode triad has a damping rate much smaller that the excitation rate of the parent mode
such that $|Q_B|\gg |Q_A|$ or $|Q_C|\gg |Q_A|$ (see equation (31)),
the amplitude $|\delta L_{\rm rad}/L_{\rm rad}|$ of the daughter mode can be comparable to or even larger than 
that of the parent mode.
In fact, although the upper limit of $|\delta L_{\rm rad}/L_{\rm rad}|$
for the parent modes is of order $\sim 10^{-4}$, 
there are many daughter modes whose amplitude is as large as $|\delta L_{\rm rad}/L_{\rm rad}|\gtsim 10^{-3}$.
Since $|\bar\omega_B|+|\bar\omega_C|\ltsim |\bar\omega_A|$, the daughter modes are likely to be in the low frequency domain of $|\bar\sigma|$
for the case of $\bar\Omega=0.01$ since the term $m\bar\Omega$ in $\bar\sigma=\bar\omega-m\bar\Omega$ is small compared to $\bar\omega$
and hence $\bar\sigma\sim\bar\omega$.
For the case of $\bar\Omega=0.2$, however, the term $m\bar\Omega$ can be comparable to $\bar\omega$ and even the daughter modes
can have the inertial frame oscillation frequencies comparable to those of the parent modes.
The range of the fractional amplitude $|\delta L_{\rm rad}/L_{\rm rad}|$ for $\bar\Omega=0.2$ is much wider than
that for $\bar\Omega=0.01$, although
the upper limits are almost the same.
It may be interesting to note that for the case of $\bar\Omega=0.2$, 
low radial order odd $l^\prime=m=1$ $r$-modes, which are linearly stable but non-linearly excited,
have very low oscillation frequency $\bar\sigma\sim 0$ in the inertial frame, although the amplitudes
are not very high.
As indicated by the existence of vertical sequences of open symbols (daughter modes) in the panels (a) and (b), 
there arise some cases in which a stable low radial order $r$-mode or $g$-mode, which has a very small damping rate $\bar\gamma$, 
is shared by several parent modes, indicating that one stable daughter mode has different amplitudes depending on the mode triads
it belongs to.
This may suggest that the set of daughter modes we use for the computation of $\eta_{ABC}$ is not large enough, or
that the three mode non-linear coupling theory is
too simplified to be applied to the case where modes having an extremely small damping rate exist
in dense frequency spectra of oscillation modes.

\section{conclusions}

Using the weakly non-linear theory of oscillation,
we have estimated the amplitudes of low $m$ $g$-modes and $r$-modes in a slightly evolved $4M_\odot$ main sequence model, 
assuming the mode amplitudes are limited by parametric instability between one unstable mode and two stable modes.
Here, the unstable low frequency modes are assumed destabilized by the $\kappa$-mechanism associated with the iron opacity bump, 
and we have taken account of 
the effects of rotation on low frequency modes in the traditional approximation.
For a given unstable mode (parent mode A), we compute three mode non-linear coupling coefficient $\eta_{ABC}$
for various combinations of two stable modes (daughter modes B and C), and we choose, among the numerous combinations, the one that gives the smallest critical amplitude.
The traditional approximation is employed in order to reduce the amount of computing time
necessary to find the optimal combination of daughter modes.
It is important to note that since we can use only a limited set of daughter modes the critical amplitude thus determined for a parent mode
should be regarded as an upper limit.
The critical amplitude essentially depends on $|\eta_{ABC}|$ and $\sqrt{Q_BQ_C}$ if we assume resonant mode coupling satisfying $\Delta\omega\sim0$,
and the smallest critical amplitude may take place when either $|\eta_{ABC}|$ or $\sqrt{Q_BQ_C}$ is very large or when both of them are large.
If the damping rate of a parametrically excited daughter mode in a mode triad is less than the growth rate of
the parent mode, the equilibrium amplitude of the daughter mode can be 
larger than the parent mode.
It is therefore likely that parametrically destabilized daughter modes like low radial order $g$-modes and $r$-modes
are among the periodicities observed in a rapidly rotating B star.
The fractional amplitudes $|\delta L_{\rm rad}/L_{\rm rad}|$ of the parent and daughter modes can be of order 
$\sim10^{-4}$ to $\sim 10^{-3}$ for the main sequence model, the magnitudes of which may be 
consistent with those observed in B type variable stars (e.g., Huat et al 2009; Diago et al 2009;
Neiner et al 2009; Cameron et al 2008; Balona et al 2011).
We also find that the amplitudes $|\delta L_{\rm rad}/L_{\rm rad}|$ of the parent mode
tend to be large for high $l$ values, 
although the visibility of the modes decreases with increasing $l$.

We find that $r$-modes significantly affects the amplitude determination of low frequency modes in a rapidly rotating star.
Since low radial order $r$-modes of the model
have damping rates $\bar\gamma$ as small as or even smaller than $\sim 10^{-10}$, 
the mode triads giving the smallest critical amplitude for the low frequency parent modes are likely to have
a low radial order $r$-mode as a member.
When a mode triad has a low radial order $r$-mode, the equilibrium amplitude of the parent mode tends to be smaller
than those for the mode triads without $r$-modes.
We find some cases in which a low radial order $r$-mode is shared by several parent modes.
We think the degeneracy of the daughter mode in mode triads is a problem in the weak non-linear coupling theory we use,
since the equilibrium amplitude of the daughter mode depends on the parent modes it is coupled to.
This degeneracy might be removed if we use an enlarged set of daughter modes to determine the optimal combination, or
this degeneracy may suggest that the three mode coupling theory we use is too simplified to be applied to the problem we are considering, that is, 
we have to consider higher order non-linear mode coupling to lift the degeneracy.
In spite of the problem,
the weakly non-linear theory could be useful when we try to compare theoretical mode calculations to observations.
Generally, the number of observationally detected low frequency modes for SPB stars
is much smaller than that of theoretically calculated linearly unstable modes
(e.g., Walker et al 2005, Saio et al 2007), and we may use
the weakly non-linear theory to decide which linearly unstable modes can have amplitudes
large enough to be detected observationally.
We may suggest that, if a low radial order $l^\prime=m=1$ $r$-mode, which may be linearly stable, 
is parametrically excited, the $r$-mode can produce
very long period variations observed in an inertial frame.
We may attribute very slow pulsations detected
in SPBe stars (e.g., Walker et al 2005, Saio et al 2007) to low radial order $l^\prime=m=1$ $r$-modes, although the amplitudes would not be very high as indicated by Figure 6.

We have carried out an additional calculation to obtain the optimal critical amplitudes $|c_{A:c}|$ 
for the parent modes by extending the set of daughter modes from $l_{\rm max}=|m|+1=6$ to
$l_{\rm max}=8$ for $\bar\Omega=0.01$, and we obtained almost the same result for their amplitudes $|\delta L_{\rm rad}/L_{\rm rad}|$.
However, it is extremely time consuming to carry out similar calculations for $l_{\rm max}$ much larger than $l_{\rm max}\sim 10$, and
from the numerical results we currently have it would be fair to say 
we are not able to correctly specify what the most likely degrees of the daughter modes are for the parent modes.
To construct a set of daughter modes extended for a very large $l_{\rm max}$, 
asymptotic methods would be useful to represent the eigenfunctions and eigenfrequencies for high $l$ and high radial order $g$-modes and to
calculate the coupling coefficient $\eta_{ABC}$ (e.g., Dziembowski 1982), and we may use an asymptotic treatment
by Lee \& Saio (1989) for low frequency modes in uniformly rotating stars.
Extending our weakly non-linear analysis to large values of $l$,
we can also consider non-linear 
couplings not only between a parent mode and many pairs of daughter modes having similar frequencies but also between a daughter mode and granddaughter
modes with frequencies still lower than that of the daughter mode (e.g., Kumar \& Goodman 1996).
The problems of these highly multiple mode couplings could be important for the amplitude determination for both the parent modes and daughter modes,
and we may have to include these mechanisms in our analysis to obtain definite answers for the oscillation amplitudes in the stars.

We have assumed uniform rotation in the present analysis.
It is, however, quite likely that differential rotation is a rule
in reality in a rotating star, and that
even a weak differential rotation would affect the frequency spectrum of the
low frequency modes.
We need to understand the property of low frequency modes in a differentially
rotating star as well as how a differential rotation law is established in a star.
The problem of differential rotation in a rotating star is quite difficult to find answer and is beyond the scope of this paper.
Since we used the traditional approximation to compute low frequency modes in a rotating star,
we could not correctly take account of the effects of linear coupling between the modes associated with different $\lambda_{km}$s.
As discussed by Aprilia et al (2011), the linear coupling between low frequency modes
tends to preferentially stabilize retrograde $g$-modes for rapidly rotating B stars, particularly 
for those having lower effective temperatures.
It is therefore desirable to
use the expansion method (e.g., Lee \& Saio 1987; Lee \& Baraffe 1995) to compute low frequency modes in a rotating star 
for the weakly non-linear coupling calculation (as well as for the analysis of the pulsational stability),
although the calculation using the expansion method would be much more time consuming than that using the traditional approximation to find the optimal
combination of daughter modes for a given parent mode.
In a B type main sequence star, 
a $\mu$-gradient zone with a high Brunt V\"ais\"al\"a frequency forms above the convective core as it evolves from the ZAMS,
and the $\mu$-gradient zone has the effect of enhancing the coupling coefficient $|\eta_{ABC}|$
compared to the case of the ZAMS model with no $\mu$-gradient zone.
This enhancement of $|\eta_{ABC}|$ would affect the equilibrium amplitudes $|c_{A:e}|$ of the low frequency modes.
It is therefore important to examine the effects of stellar evolution on the quantities $|\eta_{ABC}|$ and $|c_{A:e}|$ and hence on the amplitude determination
of the low frequency modes in SPB stars in the weakly non-linear coupling theory.

\begin{appendix}

\section{Oscillation Equation in rotating stars in the traditional approximation}

The oscillation equations for uniformly rotating stars in the traditional approximation may be given by (e.g., Lee \& Saio 1990)
\be
r{dz_1\over dr}=\left({V\over\Gamma_1}-3\right)z_1+\left({\lambda_{km}\over c_1\bar\omega^2}-{V\over\Gamma_1}\right) z_2,
\ee
\be
r{dz_2\over dr}=\left(c_1\bar\omega^2+rA\right)z_1+\left(1-U-rA\right)z_2,
\ee
where 
\be
z_1={\xi^r(r)\over r}, \quad z_2={p^\prime(r)\over \rho g r},
\ee
and
\be
V=-{d\ln p\over d\ln r}, \quad U={d\ln M_r\over d\ln r}, \quad rA={d\ln\rho\over d\ln r}-{1\over\Gamma_1}{d\ln p\over d\ln r}, 
\ee
$
c_1={(r/R)^3/ (M_r/M)}, 
$
$M_r=\int_0^R4\pi r^2\rho dr$, $g=GM_r/r^2$, $\Gamma_1=\left(\partial\ln p/\partial\ln\rho\right)_{\rm ad}$, 
and $M$ and $R$ are the mass and radius of the star, and $G$ is the gravitational constant.
With appropriate boundary conditions imposed at the centre and surface of the star, we solve the above set of 
differential equations as a boundary-eigenvalue problem for the eigenfrequency $\omega$.
Since $\xi^r=rz_1\tilde\Theta_{km}$, $\xi^\theta=rz_2\tilde\Theta^\theta_{km}/c_1\bar\omega^2$, and 
$\xi^\phi=r z_2\tilde\Theta^\phi_{km}/c_1\bar\omega^2$, the derivatives $\partial\xi^r/\partial r$,
$\partial\xi^\theta/\partial r$, and
$\partial \xi^\phi/\partial r$ may be calculated by making use of equations (A1) and (A2).

Using the eigenfunctions $\pmb{\xi}$, the oscillation energy $\epsilon$ observed in the corotating frame 
of the star is given by
(Lee \& Saio 1990)
\be
\epsilon\equiv\omega b=\omega^2\int_0^R\pmb{\xi}^*\cdot\pmb{\xi}\rho r^2dr,
\ee
and it is interesting to note that for positive $\lambda_{km}$
\be
\left(\lambda_{km}+\nu{\partial\lambda_{km}\over\partial \nu}\right)\int_{-1}^1d\mu\left|\Theta_{km}\right|^2=
\int_{-1}^1d\mu\left(\left|\Theta^\theta_{km}\right|^2+\left|\Theta^\phi_{km}\right|^2\right).
\ee

\section{Calculation of Coupling Coefficient $\kappa_{ABC}$}

If we neglect the boundary terms using the pressure zero surface boundary condition,
by use of partial integrations
we can rewrite the expression for the coupling coefficient $\kappa_{ABC}$ as (Schenk et al 2002)
\be
\kappa_{ABC}=\kappa_{ABC}^{(1)}+\kappa_{ABC}^{(2)}+\kappa_{ABC}^{(3)}+\kappa_{ABC}^{(4)},
\ee
where
\be
\kappa_{ABC}^{(1)}={1\over 2}\int d^3\pmb{x}p\left(\Gamma_1-1\right)\left(\Xi_{AB}\nabla\cdot\pmb{\xi}_C
+\Xi_{BC}\nabla\cdot\pmb{\xi}_A+\Xi_{CA}\nabla\cdot\pmb{\xi}_B\right),
\ee
\be
\kappa_{ABC}^{(2)}=
{1\over 2}\int d^3\pmb{x}p\left[\left(\Gamma_1-1\right)^2+{\partial\Gamma_1\over\partial\ln\rho}\right]\nabla\cdot\pmb{\xi}_A\nabla\cdot\pmb{\xi}_B\nabla\cdot\pmb{\xi}_C,
\ee
\be
\kappa_{ABC}^{(3)}
={1\over 2}\int d^3\pmb{x}p\left(\chi_{ABC}+\chi_{ACB}\right),
\ee
\be
\kappa_{ABC}^{(4)}=
-{1\over 2}\int d^3\pmb{x}\rho\xi^i_A\xi^j_B\xi^k_C\nabla_i\nabla_j\nabla_k\Phi,
\ee
where 
$
\Xi_{AB}=\delta^i_j\Xi_i^j\left(\pmb{\xi}_A,\pmb{\xi}_B\right), 
$
$
\chi_{ABC}=\delta_i^j\chi^i_j\left(\pmb{\xi}_A,\pmb{\xi}_B,\pmb{\xi}_C\right),
$
$
\chi^i_j\left(\pmb{\xi}_A,\pmb{\xi}_B,\pmb{\xi}_C\right)=(\nabla_l\xi^i_A)(\nabla_k\xi^l_B)(\nabla_j\xi^k_C),
$
and the repeated indices imply the summation over the indices from 1 to 3.
We note that since $\chi_{ABC}=\chi_{BCA}=\chi_{CAB}$ and $\chi_{ACB}=\chi_{BAC}=\chi_{CBA}$, 
the coefficient $\kappa_{ABC}$ does not depend on the order of the indices $A$, $B$, and $C$.
Note that if we consider $\kappa_{ABC}(r)=\int_0^r\left(d\kappa_{ABC}/dr\right)dr$, we have $\kappa_{ABC}=\kappa_{ABC}(R)$.

In spherical polar coordinates $(r,\theta,\phi)$, the covariant derivatives of the displacement vectors are
\be
\xi^r_{;r}={\partial \xi^r\over\partial r}={\partial (rz_1)\over\partial r}\tilde\Theta,
\ee
\be
\xi^r_{;\theta}={\partial\xi^r\over\partial\theta}-\xi^\theta=r\left(z_1\partial\tilde\Theta-{z_2\over c_1\bar\omega^2}\tilde\Theta^\theta\right),
\ee
\be
\xi^r_{;\phi}={\partial\xi^r\over\partial\phi}-\sin\theta\xi^\phi=\rmi r\left(z_1m\tilde\Theta
-{z_2\over c_1\bar\omega^2}\sin\theta\tilde\Theta^\phi\right),
\ee
\be
\xi^\theta_{;r}={1\over r}{\partial\xi^\theta\over \partial r}={1\over r}{\partial\over \partial r}\left(r{z_2\over c_1\bar\omega^2}\right)\tilde\Theta^\theta,
\ee
\be
\xi^\theta_{;\theta}={1\over r}{\partial\xi^\theta\over\partial\theta}+{\xi^r\over r}=
z_1\tilde\Theta+{z_2\over c_1\bar\omega^2}\partial\tilde\Theta^\theta,
\ee
\be
\xi^\theta_{;\phi}={1\over r}{\partial\xi^\theta\over\partial\phi}-{1\over r}\cos\theta\xi^\phi
=\rmi {z_2\over c_1\bar\omega^2}\tilde\Theta^\theta_\phi,
\ee
\be
\xi^\phi_{;r}={1\over r\sin\theta}{\partial\xi^\phi\over\partial r}
=\rmi r{\partial\over \partial r}\left(r{z_2\over c_1\bar\omega^2}\right){\tilde\Theta^\phi\over \sin\theta},
\ee
\be
\xi^\phi_{;\theta}={1\over r\sin\theta}{\partial\xi^\phi\over\partial\theta}
=\rmi {z_2\over c_1\bar\omega^2}{\partial\tilde\Theta^\phi\over\sin\theta},
\ee
\be
\xi^\phi_{;\phi}={1\over r\sin\theta}{\partial\xi^\phi\over\partial\phi}+{\xi^r\over r}
+{1\over r}{\cos\theta\over\sin\theta}\xi^\theta=z_1\tilde\Theta+{z_2\over c_1\bar\omega^2}\tilde\Theta^\phi_\phi,
\ee
where we have written $\xi^i_{;j}$, instead of $\nabla_j\xi^i$, for the covariant derivatives, and
\be
\partial\tilde\Theta={\partial\tilde\Theta\over\partial\theta}, \quad
\partial\tilde\Theta^\theta={\partial\tilde\Theta^\theta\over\partial\theta}, \quad
\partial\tilde\Theta^\phi={\partial\tilde\Theta^\phi\over\partial\theta},
\ee
and
\be
\tilde\Theta^\theta_\phi=m\tilde\Theta^\theta-\cos\theta\tilde\Theta^\phi, \quad
\tilde\Theta^\phi_\phi=-{m\over\sin\theta}\tilde\Theta^\phi+{\cos\theta\over\sin\theta}\tilde\Theta^\theta.
\ee
Note that we have omitted the subscript $km$ attached to the functions $\tilde\Theta_{km}$, $\tilde\Theta^\theta_{km}$, $\tilde\Theta^\phi_{km}$, etc., for simplicity.

The integrands of $\kappa$ may be given by
\be
\begin{array}{l}
\displaystyle
\Xi_{AB}\nabla\cdot\pmb{\xi}_C
+\Xi_{BC}\nabla\cdot\pmb{\xi}_A+\Xi_{CA}\nabla\cdot\pmb{\xi}_B=
{1\over 2}S\left({\partial(rz_1)\over \partial r}{\partial(rz_1)\over \partial r}H:\tilde\Theta\tilde\Theta\tilde\Theta\right)
\\
\displaystyle
+{1\over 2}S\left({z_2\over c_1\bar\omega^2}{z_2\over c_1\bar\omega^2}H:\left[\partial\tilde\Theta^\theta
\partial\tilde\Theta^\theta+\tilde\Theta^\phi_\phi\tilde\Theta^\phi_\phi
-2{\tilde\Theta^\theta_\phi\partial\tilde\Theta^\phi\over\sin\theta}\right]\tilde\Theta\right)
+S\left(z_1z_1H:\tilde\Theta\tilde\Theta\tilde\Theta\right)
+S\left({z_2\over c_1\bar\omega^2}z_1H:\left[\partial\tilde\Theta^\theta\tilde\Theta+\tilde\Theta^\phi_\phi\tilde\Theta\right]\tilde\Theta\right)
\\
\displaystyle
+S\left(z_1\left({\partial\over\partial r}r{z_2\over c_1\bar\omega^2}\right)H:\left[\partial\tilde\Theta\tilde\Theta^\theta-{m\tilde\Theta\tilde\Theta^\phi\over\sin\theta}\right]\tilde\Theta\right)
-S\left(H{z_2\over c_1\bar\omega^2}\left({\partial\over\partial r}r{z_2\over c_1\bar\omega^2}\right):
\tilde\Theta\left[\tilde\Theta^\theta\tilde\Theta^\theta-\tilde\Theta^\phi\tilde\Theta^\phi\right]\right),\\
\end{array}
\ee
\be
\nabla\cdot\pmb{\xi}_A\nabla\cdot\pmb{\xi}_B\nabla\cdot\pmb{\xi}_C=H_AH_BH_C\tilde\Theta_A\tilde\Theta_B\tilde\Theta_C,
\ee
\be
\begin{array}{l}
\displaystyle
\chi_{ABC}+\chi_{ACB}=S\left({\partial(rz_1)\over\partial r}\left(z_1-{z_2\over c_1\bar\omega^2}\right)
\left({\partial\over\partial r}r{z_2\over c_1\bar\omega^2}\right):\tilde\Theta\left[\tilde\Theta^\theta\tilde\Theta^\theta-\tilde\Theta^\phi\tilde\Theta^\phi\right]\right)\\
\displaystyle
+S\left({\partial(rz_1)\over\partial r}z_1
\left({\partial\over\partial r}r{z_2\over c_1\bar\omega^2}\right):\tilde\Theta\left(\partial\tilde\Theta\tilde\Theta^\theta-{m\tilde\Theta\tilde\Theta^\phi\over\sin\theta}\right)-\tilde\Theta\left[\tilde\Theta^\theta\tilde\Theta^\theta-\tilde\Theta^\phi\tilde\Theta^\phi\right]\right)\\
\displaystyle
+S\left(z_1{z_2\over c_1\bar\omega^2}\left({\partial\over\partial r}r{z_2\over c_1\bar\omega^2}\right):
\partial\tilde\Theta\partial\tilde\Theta^\theta\tilde\Theta^\theta-{m\tilde\Theta\tilde\Theta^\phi_\phi
\tilde\Theta^\phi+m\tilde\Theta\partial\tilde\Theta^\phi\tilde\Theta^\theta+\partial\tilde\Theta\tilde\Theta^\theta_\phi\tilde\Theta^\phi\over\sin\theta}
-\tilde\Theta\left[\tilde\Theta^\theta\tilde\Theta^\theta-\tilde\Theta^\phi\tilde\Theta^\phi\right]\right)\\
\displaystyle
+S\left({z_2\over c_1\bar\omega^2}{z_2\over c_1\bar\omega^2}\left({\partial\over\partial r}r{z_2\over c_1\bar\omega^2}\right):\tilde\Theta^\phi\tilde\Theta^\phi_\phi\tilde\Theta^\phi-\tilde\Theta^\theta\partial\tilde\Theta^\theta\tilde\Theta^\theta+\tilde\Theta^\phi\partial\tilde\Theta^\phi\tilde\Theta^\theta+{\tilde\Theta^\theta_\phi\tilde\Theta^\theta\tilde\Theta^\phi\over\sin\theta}\right)\\
\displaystyle
+S\left(z_1z_1\left({\partial\over\partial r}r{z_2\over c_1\bar\omega^2}\right):\partial\tilde\Theta\tilde\Theta\tilde\Theta^\theta-{m\tilde\Theta\tilde\Theta\tilde\Theta^\phi\over\sin\theta}\right)\\
\displaystyle
-S\left({z_2\over c_1\bar\omega^2}{z_2\over c_1\bar\omega^2}{z_2\over c_1\bar\omega^2}:
{\partial\tilde\Theta^\theta\tilde\Theta^\theta_\phi\partial\tilde\Theta^\phi+\partial\tilde\Theta^\phi\tilde\Theta^\theta_\phi\tilde\Theta^\phi_\phi\over\sin\theta}\right)
-2S\left(z_1{z_2\over c_1\bar\omega^2}{z_2\over c_1\bar\omega^2}:{\tilde\Theta\tilde\Theta^\theta_\phi\partial\tilde\Theta^\phi\over\sin\theta}\right)\\
\displaystyle
+{1\over 2}S\left(z_1{z_2\over c_1\bar\omega^2}{z_2\over c_1\bar\omega^2}:\tilde\Theta\tilde\Theta^\phi_\phi\tilde\Theta^\phi_\phi+\tilde\Theta\partial\tilde\Theta^\theta\partial\tilde\Theta^\theta\right)
+{1\over 2}S\left(z_1z_1{z_2\over c_1\bar\omega^2}:\tilde\Theta\tilde\Theta\partial\tilde\Theta^\theta+
\tilde\Theta\tilde\Theta\tilde\Theta^\phi_\phi\right)\\
\displaystyle
+\left({\partial(rz_1)_A\over\partial r}{\partial(rz_1)_B\over\partial r}{\partial(rz_1)_C\over\partial r}
+2(z_{1})_A(z_{1})_B(z_{1})_C\right)\tilde\Theta_A\tilde\Theta_B\tilde\Theta_C\\
\displaystyle
+\left({z_2\over c_1\bar\omega^2}\right)_A\left({z_2\over c_1\bar\omega^2}\right)_B\left({z_2\over c_1\bar\omega^2}\right)_C
\left(\tilde\Theta^\phi_{\phi A}\tilde\Theta^\phi_{\phi B}\tilde\Theta^\phi_{\phi C}
+\partial\tilde\Theta^\theta_A\partial\tilde\Theta^\theta_B\partial\tilde\Theta^\theta_C
\right),
\end{array}
\ee
and
\be
r^{-3}\xi^i_A\xi^j_B\xi^k_C\Phi_{;ijk}={1\over 2}S\left(z_1{z_2\over c_1\bar\omega_2}{z_2\over c_1\bar\omega_2}:
\tilde\Theta\left[\tilde\Theta^\theta\tilde\Theta^\theta-\tilde\Theta^\phi\tilde\Theta^\phi\right]\right){\partial\over\partial r}\left({1\over r}{\partial\Phi\over\partial r}\right)
+(z_1)_A(z_1)_B(z_1)_C\tilde\Theta_A\tilde\Theta_B\tilde\Theta_C{\partial^3\Phi\over\partial r^3},
\ee
where the function $H(r)$ is defined by
\be
\nabla\cdot\pmb{\xi}=H(r)\tilde\Theta(\theta,\phi)=-{V\over\Gamma_1}\left(z_2-z_1\right)\tilde\Theta(\theta,\phi),
\ee
and
\be
\begin{array}{r}
\displaystyle
S\left(f^1f^2f^3:p^1p^2p^3\right)=
f^1_Af^2_Bf^3_Cp^1_Ap^2_Bp^3_C
+f^1_Af^2_Cf^3_Bp^1_Ap^2_Cp^3_B
+f^1_Bf^2_Cf^3_Ap^1_Bp^2_Cp^3_A\\
\displaystyle
+f^1_Bf^2_Af^3_Cp^1_Bp^2_Ap^3_C
+f^1_Cf^2_Af^3_Bp^1_Cp^2_Ap^3_B
+f^1_Cf^2_Bf^3_Ap^1_Cp^2_Bp^3_A,
\end{array}
\ee
where the functions $f^j$s depend only on $r$ and the functions $p^j$s only on $\theta$ and $\phi$.
Note that 
\be
S\left(f^1f^2f^3:p^1p^2p^3\right)=S\left(f^2f^1f^3:p^2p^1p^3\right)=S\left(f^1f^3f^2:p^1p^3p^2\right)=S\left(f^3f^2f^1:p^3p^2p^1\right)=\cdots.
\ee
Integrating $S\left(f^1f^2f^3:p^1p^2p^3\right)$ over a sphere of radius $r$, we obtain
\be
\begin{array}{r}
\displaystyle
\int S\left(f^1f^2f^3:p^1p^2p^3\right)d\Omega=f^1_Af^2_Bf^3_CZ^{123}_{ABC}
+f^1_Af^2_Cf^3_BZ^{123}_{ACB}+f^1_Bf^2_Cf^3_AZ^{123}_{BCA}~\\
+f^1_Bf^2_Af^3_CZ^{123}_{BAC}
+f^1_Cf^2_Af^3_BZ^{123}_{CAB}+f^1_Cf^2_Bf^3_AZ^{123}_{CBA},
\end{array}
\ee
where
\be
Z^{123}_{ABC}=\int p^1_Ap^2_Bp^3_C d\Omega.
\ee

\end{appendix}

\end{document}